\documentclass[preprint]{revtex4}
\usepackage[pdftex]{graphicx}

\usepackage{amssymb,amsfonts,amsmath}

\newcommand{\vecb}[1]{\mbox{\boldmath$#1$}}
  
\newcommand{\um}{$\mu$m}
\newcommand{\figFingerprintingSchematics}{aom_schematics_lens_2_copy_nb.pdf}
\newcommand{\figDiffLim}{difflim1_bw.pdf}

\newcommand{\figFingerprintingTwoP}{figureFingerprExpt.pdf}
\newcommand{\figBragg}{bragg_imagingBlue.pdf}
\begin{document}

%%%%%%%%%%%%%%%%%%%%%%%%%%%%%%

%% For titles, only capitalize the first letter
%% \title{Almost sharp fronts for the surface quasi-geostrophic equation}

\title{Super-resolution imaging using spatial Fourier transform infrared spectroscopy}

%% Enter authors via the \author command.  
%% Use \affil to define affiliations.
%% (Leave no spaces between author name and \affil command)

%% Note that the \thanks{} command has been disabled in favor of
%% a generic, reserved space for PNAS publication footnotes.

%% \author{<author name>
%% \affil{<number>}{<Institution>}} One number for each institution.
%% The same number should be used for authors that
%% are affiliated with the same institution, after the first time
%% only the number is needed, ie, \affil{number}{text}, \affil{number}{}
%% Then, before last author ...
%% \and
%% \author{<author name>
%% \affil{<number>}{}}

%% For example, assuming Garcia and Sonnery are both affiliated with
%% Universidad de Murcia:
%% \author{Roberta Graff\affil{1}{University of Cambridge, Cambridge,
%% United Kingdom},
%% Javier de Ruiz Garcia\affil{2}{Universidad de Murcia, Bioquimica y Biologia
%% Molecular, Murcia, Spain}, \and Franklin Sonnery\affil{2}{}}

\author{Leonid Alekseyev}
\affiliation{Electrical Engineering Department,
  Princeton University, Princeton, NJ  08544}
\affiliation{Electrical
  Engineering Department, Purdue University, West Lafayette, IN  47906}
\author{Evgenii Narimanov}
\affiliation{Electrical
  Engineering Department, Purdue University, West Lafayette, IN  47906}
\author{Jacob Khurgin}
\affiliation{Electrical Engineering Department, Johns Hopkins 
  University, Baltimore, MD  21218}

%% The \maketitle command is necessary to build the title page.

%%%%%%%%%%%%%%%%%%%%%%%%%%%%%%%%%%%%%%%%%%%%%%%%%%%%%%%%%%%%%%%%

\begin{abstract}
Spatial resolution of most imaging devices is fundamentally restricted
by diffraction.  This limitation is manifested in the
loss of high spatial frequency information contained in evanescent
waves.  As a result, conventional far-field optics yields no
information about an object's subwavelength features.  Here we propose
a novel approach to recovering evanescent waves in the far field,
thereby enabling subwavelength-resolved imaging and spatial
spectroscopy.  Our approach relies on shifting the frequency and the
wave vector of near-field components via scattering on acoustic
phonons.  This process effectively removes the spatial frequency
cut-off for unambiguous far field detection.  A straightforward
extension of this technique, which we call spatial Fourier transform
infrared spectroscopy, allows to preserve phase information,
making it possible to perform 3D subwavelength imaging.  We discuss
the implementation of such a system in the mid-IR and THz bands, with
possible extension to other spectral regions.
\end{abstract}

\maketitle

%% When adding keywords, separate each term with a straight line: |
% \keywords{superresolution | spectroscopy | detection}

%% Optional for entering abbreviations, separate the abbreviation from
%% its definition with a comma, separate each pair with a semicolon:
%% for example:
%% \abbreviations{SAM, self-assembled monolayer; OTS,
%% octadecyltrichlorosilane}

%\abbreviations{S-FTIR, spatial Fourier transform infrared spectroscopy}

%% The first letter of the article should be drop cap: \dropcap{}
%\dropcap{I}n this article we study the evolution of ''almost-sharp'' fronts

%% Enter the text of your article beginning here and ending before
%% \begin{acknowledgements}
%% Section head commands for your reference:
%% \section{}
%% \subsection{}
%% \subsubsection{}

\begin{figure*}[b]
\centerline{\scalebox{1.481}{\includegraphics{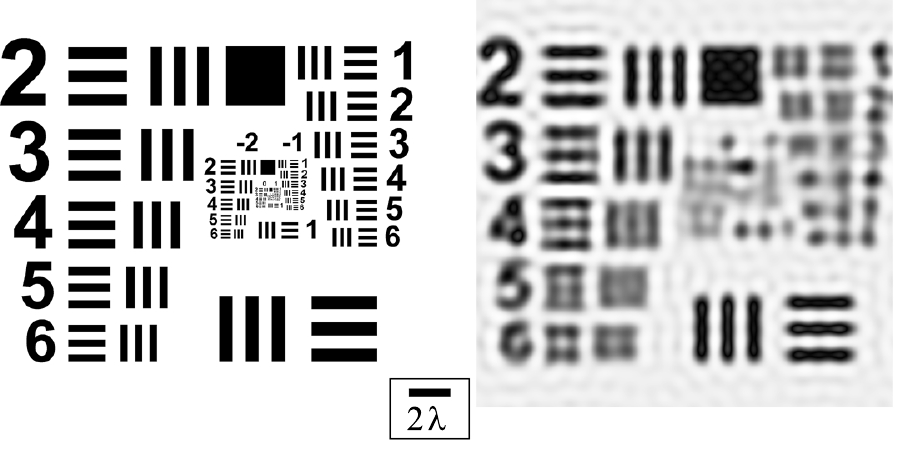}}}
\caption{Computed effects of the diffraction limit on an optical test
  target (assuming coherent illumination).  The target consists of
  labeled sets of line groups, which halve in size every six elements.
  The last resolvable set of lines (group 5, left column) corresponds
  to 1.5 lines/$\lambda$.} \label{fig:diffLim}
\end{figure*}

Optical microscopy is widely used for probing the nanoscale, with
particularly important applications in the life sciences.  It is fast,
noninvasive, and can be coupled with fluorescence and spectroscopy
studies to uncover the structure, composition, and dynamics of
biological samples.  However, due to the diffraction limit much of the
important structural information about the object under study becomes
lost~\cite{Abbe1873}.  We illustrate the extent of this loss by comparing the image
[Fig.\ref{fig:diffLim}(b)] of a subwavelength optical target
[Fig.\ref{fig:diffLim}(a)] with one that would be produced by an ideal
diffraction-limited optical system (which could be implemented using
e.g.\ synthetic aperture techniques~\cite{Lukosz1967,BrueckOL1999}).
It is apparent the details smaller than $\sim\lambda/2$ cannot be
resolved.  Because of this wavelength dependence, the resolution of
the image particularly suffers in the mid-IR and THz spectral bands.
This creates a significant stumbling block for studying microscopic
chemical morphology.  For instance, mid-infrared spectroscopy -- a
crucial tool for chemical identification -- cannot be used for
structure-specific analysis of biological samples due to the inability
to resolve small cells, bacteria, and cellular subunits.

There exist certain techniques that can improve spatial resolution of
optical imaging beyond the diffraction limit.  One category of methods
involves scanning probes~\cite{Dragnea2001,Keilmann1999,Planken2002},
which rely on the detection of evanescent waves in the near field of
the sample.  Nonlinear processes form another set of approaches; these
include coherent anti-Stokes Raman spectroscopy
(CARS)~\cite{Xie1999,Ito2004}, as well as RESOLFT techniques (which
utilize saturable fluorescence transitions)~\cite{Hell2007}.  All the
above solutions typically involve complicated and expensive
experimental setups.  This limits their widespread availability, as
well as their potential for creating compact integrated detection
and/or imaging devices.  In addition, some setups may be difficult to
adapt for operation in mid-IR or THz.

In the past few years, there emerged several intriguing approaches
towards subwavelength imaging that involve using metamaterial-based
devices, which promise to be low-cost and readily adaptable for a
variety of spectral regions.  Such devices work by recovering the
evanescent waves and the concomitant information about the object's
fine structure.  A negative index ``superlens"~\cite{pendry}, for
instance, can drastically amplify the evanescent components of the
spatial spectrum while transmitting the propagating waves.  However,
the superlens exhibits exponential sensitivity to losses, limiting
practical devices to near-field
operation~\cite{PodolskiyNarimanovNSSL}.  A more recent set of
experiments involved subwavelength scatterers placed in the near field
of a source.  By diffracting off the scatterers, evanescent waves
convert into propagating waves which could then be used to gather
information about the near-field spectrum~\cite{Durant2006} or to
attain subwavelength focusing~\cite{LeroseyFink2007}.  However, in
this far-field superlens~\cite{Durant2006}, the diffracted evanescent
waves mix with the existing propagating spectrum leading to the loss
of information about the target.

%In these devices, the diffracted evanescent waves mix with the existing propagating spectrum, which doesn't allow unambiguous %recovery of spatial frequency data.

% This places significant limitations on the accessible spatial frequency bandwidth.

In the present Letter, we propose an alternative approach to far-field
spatially resolved spectroscopy, based on a device that converts
mid-IR or THz evanescent waves to propagating waves via scattering on
acoustic phonons.  These scattered and frequency-shifted waves can be
easily decoupled from the existing propagating spectrum that forms the
regular diffraction-limited image, and with minimal processing can be
used to distinguish subwavelength features.  Furthermore, the ability
to dynamically tune the period of the acoustic grating makes this
system robust, flexible, and capable of operating over a wide range of
spatial frequencies.

\begin{figure*}
\centerline{\scalebox{.550}{\includegraphics{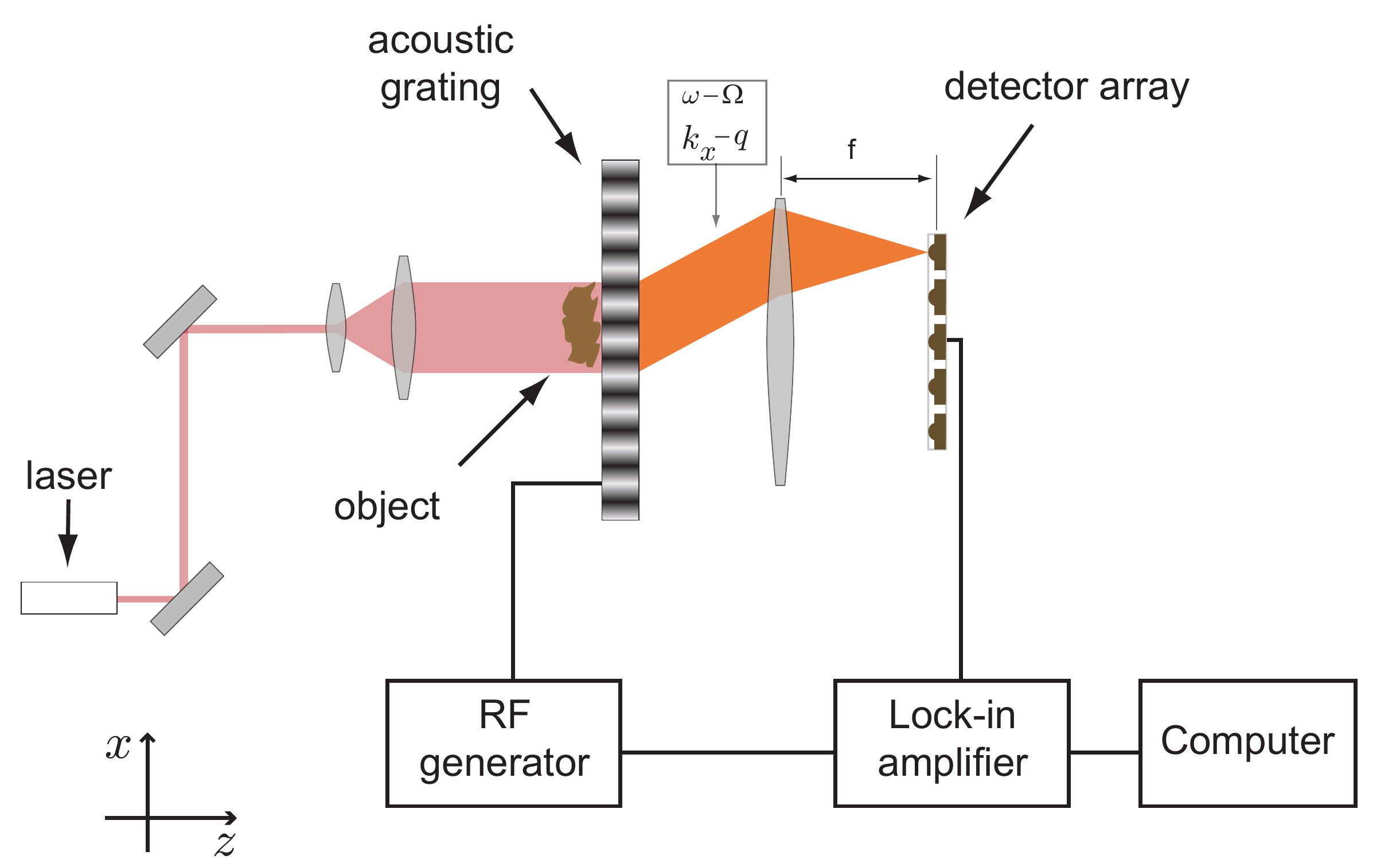}}}
\caption{Schematics of the proposed
  system. } \label{fig:fingerprintingSch}
\end{figure*}

The proposed super-resolution Fourier microscope/sensing system is
shown in Fig.~\ref{fig:fingerprintingSch}. The object is placed in the
near field of an acousto-optic modulator (AOM) and illuminated with a
plane wave from a mid-IR or THz source.  Waves scattered from the
object strike the phonon grating set up in the AOM by a running
acoustic wave with frequency~$\Omega$.  Due to scattering on the
phonons, the transverse wave vector $k_x$ of the incident radiation is
shifted by integer multiples the phonon wave vector $q$, while its
corresponding frequency is shifted by integer multiples of~$\Omega$.
For a sufficiently large $q$, the evanescent components of the
object's spatial spectrum ($|k_x| > \omega/c$) can be scattered into
the propagating waves with $|k'_x| \equiv |k_x - q| < \omega/c$.  The
various spatial frequency components can be measured using a Fourier
optics setup (e.g.\ a lens with a detector array in its focal plane).

%In order to uniquely recover the ``downshifted'' waves ($k'_x \equiv
%k_x - q$), a portion of the illuminating radiation is %shifted in
%frequency by $\Omega_b$ using a second AOM and is projected onto the
%detector.  The resulting interference produces %a beat note
%photocurrent of frequency $\Omega+\Omega_b$, which can be retrieved
%using lock-in techniques.  This signal carries the information about
%the high spatial frequency components of the object.

The amplitudes of the frequency-shifted waves can be obtained using
the standard methods of acuostooptics~\cite{KorpelBook} (cf. {\em
  Methods}).  We define $\tilde{A}_\pm = t^\pm E_{\rm in}(k_x \mp q)$,
$\tilde{A}_0 = t_0 E_{\rm in}(k_x)$ (with linear coefficients $t_0$,
$t^\pm$ describing the generation of phonon-scattered and/or device
transmission characteristics) and assume $\tilde{A}_i \gg \tilde{A}_0
\gg \tilde{A}^\pm$, where $\tilde{A}_i$ is the detected amplitude of
the illuminating wave $A_i e^{i k_0 z}$.  Averaging out the signal
over the finite detector aperture and subtracting the background
(which can be done electronically), we can write the intensity
detected by the system of Fig.~\ref{fig:fingerprintingSch} as

\begin{equation}\label{eq:measurement}
I_{\rm out} = |\tilde{A}_0|^2 + 2
\left(|\tilde{A}_i\tilde{A}_-|^2+|\tilde{A}_i\tilde{A}_+|^2\right)^{1/2}
\cos(\Omega t+ \gamma).
\end{equation}
The two terms in this equation can be decoupled using standard
techniques: the DC term is isolated with the aid of a low-pass filter,
while the term oscillating at the acoustic frequency $\Omega$ is
recoverable using lock-in detection.  For any given $k_x$, this second
term contains contributions from both $\tilde{A}_+ = t^+ E_{\rm
  in}(k_x-q)$ and $\tilde{A}_-=t^- E_{\rm in}(k_x+q)$. Although the
coupling between these two quantities, together with the lack of phase
information, makes it difficult to recover the spatial spectrum, the
information collected can be used in detecting subwavelength
morphological changes between different samples.

To illustrate this, we utilize Eq.~(\ref{eq:measurement}) to perform a
comparison between the standard optical target of
Fig.~\ref{fig:diffLim} and a modified target, where the label of every
6th line group has been randomly replaced.  The first replacement
corresponds to the last resolvable line group ($\lambda$/2.5 line
separation); the subsequent replacements correspond to halving the
size of the line groups ($\lambda$/5,$\ldots$, $\lambda$/40).  We
assume the measurement is performed by selecting an element of a
photodetector array in the observation plane and using two orthogonal
acoustic transducers to scan the acoustic wavevector within the range
$q_{x,y} \in [-25\,\omega/c, 25\,\omega/c]$.

\begin{figure*}
\centerline{\scalebox{.40}{\includegraphics{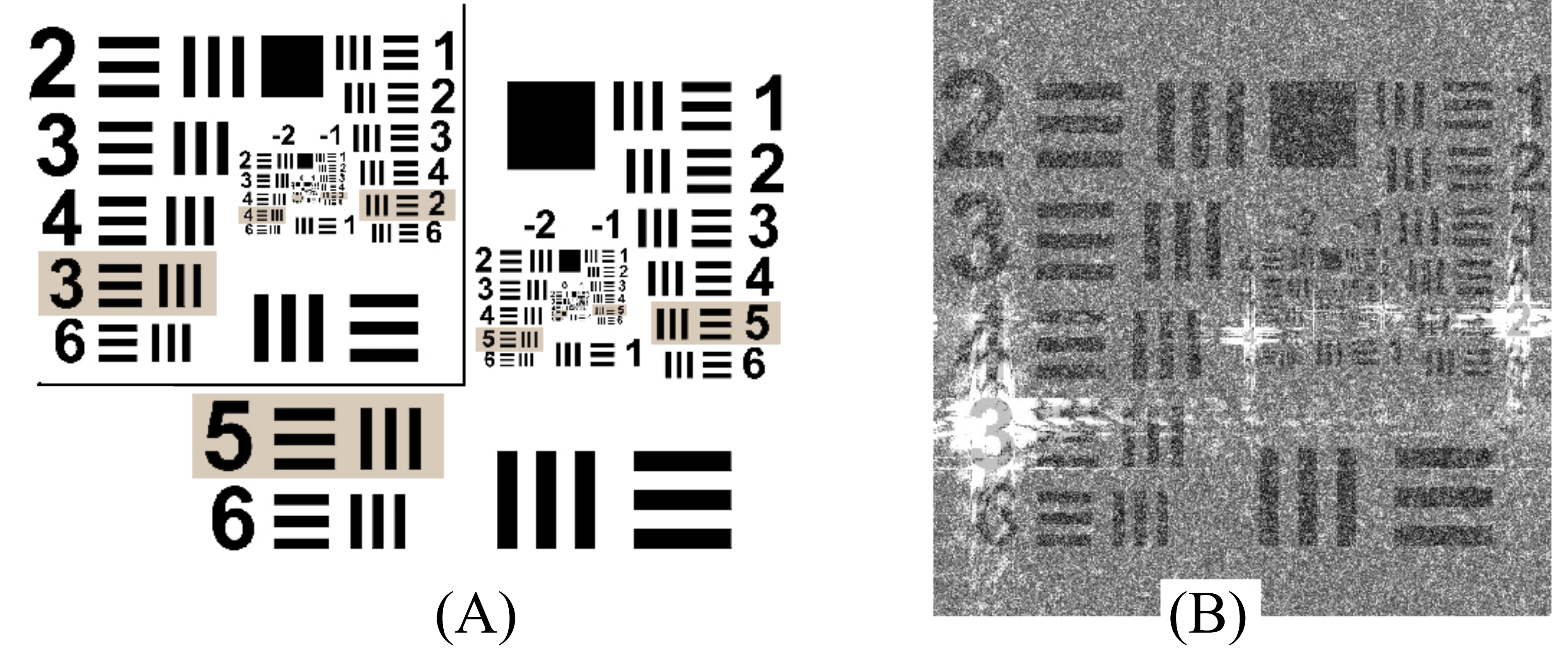}}}
\caption{(a) Optical test target and its modified version (inset).  In
  the modified target, the ``5'' label of every column has been
  replaced by another digit.  (b) Computed output of the system in the
  presence of noise (shown in grayscale) assuming a realistic, noisy
  detector with 400 active photocells.  The modified optical target is
  superimposed for illustration purposes.  The output of the system
  clearly identifies the location of every modified digit, even for
  regions far below the diffraction limit.} \label{fig:fingerprinting}
\end{figure*}

It should be emphasized that any method that relies on digital
processing of raw data can suffer form rapid -- sometimes
exponential~\cite{PodolskiyNarimanovNSSL} -- accumulation of noise.
To address this potential issue, in our computations we add a
normally-distributed random term to the AC amplitude of
Eq.~(\ref{eq:measurement}) in order to simulate noise in the system.
Because SNR is expected to be lowest for maximum values of the
acoustic wavevector $q$, we consider SNR=10 for
$q=25\,\omega/c$~\footnote{Acoustooptic diffraction efficiency, and
  hence the signal-to-noise ratio varies as $1/q$.}.  Assuming a
practical 20$\times$20 element photodetector array, we compute the
signal given by Eq.~(\ref{eq:measurement}) for the standard target, as
well as the modified target [Fig.~\ref{fig:fingerprinting}(a)].
Fig.~\ref{fig:fingerprinting}(b) shows the result of subtracting the
two datasets and performing an inverse Fourier transform, with the
resulting plot superimposed onto the modified optical target.
Evidently, every change in the original image is manifested in this
difference diagram.  Furthermore, it is largely localized in the
vicinity of the actual changed pixels.  It is possible to discern the
difference signal even from the $\lambda$/40 line group label.

The ability to distinguish between fine spatial features of optical
targets makes the system described above uniquely suited for
identifying objects based on their subwavelength spatial features.  As
a result, it may find many applications in fingerprinting and/or
detection of chemical and biological structures.

Furthermore, a straightforward modification of this setup not only
allows to measure the ``downshifted'' $\tilde{A}_-$ component
directly, but also provides a method for retrieving phase information,
making it possible to perform phase-contrast microscopy, as well as 3D
imaging on subwavelength scales.

\begin{figure*}
\centerline{\scalebox{.38}{\includegraphics{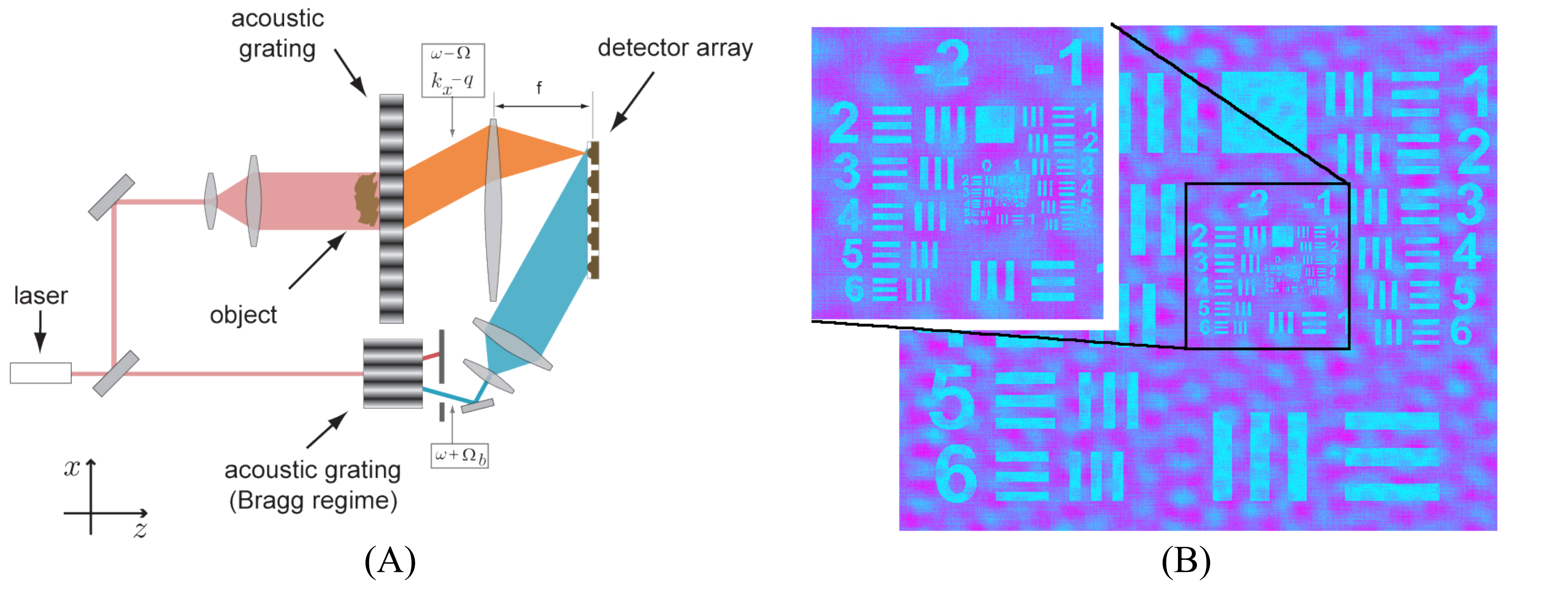}}}
\caption{(a) Schematics of the proposed system. Note the Bragg-shifted
  reference beam that aids in providing phase information.  (b) The
  computed output of the system with optical test target as the object
  in the presence of noise.} \label{fig:bragg}
\end{figure*}

  To this end, a portion of the illuminating radiation is shifted in
  frequency by $\Omega_b$ using a second AOM.  Unlike the modulator
  that interacts with light scattered from the sample in the
  Raman-Nath regime~\cite{Boyd_NLO}, this second AOM utilizes an
  appropriately oriented and longer cell to produce Bragg scattering.
  This results in a strong optical signal at frequency
  $\omega+\Omega_b$, $|\tilde{A}_b|\exp[i(k_b\cdot r - (\omega +
    \Omega_b)t)]$, which is projected onto the detector [see
    Fig.~\ref{fig:bragg}(a)].  Interference between the two optical
  signals produces beat note photocurrents with frequencies $\Omega$,
  $\Omega_b$, $\Omega_b+\Omega$, $\Omega_b-\Omega$:

\begin{align}\label{eq:iOut}
I_{\rm out}(k_x) = {} & \left|E_i \exp(i k_0 z) + 
  \tilde{A}_b\exp(i\vecb{k}\cdot\vecb{r})\right.\nonumber\\
 & + \left.[\tilde{A}^- \exp(i\Omega t) +
  \tilde{A}^+ \exp(-i\Omega t) + 
  \tilde{A}_0]\exp(i\vecb{k}\cdot\vecb{r})\right|^2 \nonumber \\ 
= {} & \ldots  + 2 |\tilde{A}^-\tilde{A}_b|
\cos[(\Omega_b+\Omega)t+\Delta\Phi^-] \nonumber\\
 & \hphantom{\ldots} + 2 |\tilde{A}^+\tilde{A}_b|
 \cos[(\Omega_b-\Omega)t+\Delta\Phi^+]+\ldots,
\end{align}
where $\Delta\Phi^\pm = (k_b-k)\cdot r - \phi^\pm$ is the phase
difference between the signal from the Bragg cell,
$|\tilde{A}_b|\exp(i k_b\cdot r)$, and the Raman-Nath-scattered signal
$\tilde{A}^\pm\exp(i k\cdot r) = |\tilde{A}^\pm|\exp[i(\phi^\pm +
  k\cdot r)]$.

Of special interest is the component at frequency $\Omega+\Omega_b$,
which carries the high spatial frequency information contained in its
modulus and its phase $\Delta\Phi^-\simeq (k_x^b-k_x)x - \phi^-$.
Both of these quantities can be retrieved using lock-in techniques.
To produce the lock-in reference, the RF signals driving the two
acoustic cells can be mixed using a nonlinear element (e.g.\ a diode)
and appropriately filtered to produce the sum frequency.  As a result,
complete information can be obtained about the complex high spatial
frequency Fourier component $\tilde{A}^-$, from which it is
straightforward to deduce the field $E_\text{in}(k_x+q)$. By
collecting data from multiple CCD pixels, as well as by varying the
acoustic wave vector $q$, information can be collected about the
spatial spectrum of the object.  The data can then be digitally
processed to produce a spatial-domain image containing subwavelength
details, as well as phase contrast.

Because the Bragg-shifted signal we use to decouple the $\tilde{A}_+$
and $\tilde{A}_-$ terms serves as a reference needed to record phase
information, and because the image is reconstructed digitally, our
technique bears some similarities with digital Fourier holography
(DFH)~\cite{OriginalGoodmanDigitalHoloPaper,HoloCCDProblem1,CucheDigitalHolo}.
However, our method contains several key enhancements over DFH.  In
conventional holography, care has to be taken to isolate the target
signal both in real and Fourier space.  This constrains reference wave
geometry, translating into limitation on the field of view, as well as
maximum attainable resolution.  The requirement that the CCD pixel
spacing must allow for imaging the reference wave fringes further
limits the resolution.  By virtue of frequency-shifting the signal, it
is possible to isolate the interference term of interest.
Furthermore, since the spatial spectrum measurements are performed not
only by selecting different CCD pixels, but also by scanning the
acoustic wavevector, the limitations of CCD's physical spatial
frequency bandwidth (introduced by pixel
granularity)~\cite{HoloCCDProblem1} can be circumvented.

We simulate the performance of the system by first using
Eq.~(\ref{eq:iOut}) to compute the response of the system to a
calibration signal having unit amplitude for all spatial frequencies.
In practice, such calibration signal might be generated by placing a
point source in the vicinity of the AOM.  Eq.~(\ref{eq:iOut}) also
provides the effective amplitude and phase transfer functions that
allow to determine the detected signal for a given input field
distribution.  Gaussian noise is added to simulate spurious signals in
the system.  The input signal can then be obtained by dividing out the
calibration quantities.  In Fig.~\ref{fig:bragg}(b) we plot the
simulated retrieved field magnitude.  Zooming in on the central part
of the test patten (figure inset) it is evident that every line group
is distinctly resolved.  Furthermore, because the phase information is
preserved, the full 3D information about the target is collected.

There exist many possible ways to enhance the functionality and the
performance of the proposed devices.  For instance, sensitivity of the
device may be improved with a subwavelength layer of highly doped
semiconductor at the front AOM facet.  When the dielectric constant of
this layer is equal to $-1$, the evanescent fields are strongly
enhanced due to resonant coupling to surface plasmons in the doped
layer (a phenomenon known as ``poor-man's
superlensing''~\cite{pendry}), leading to better SNR at the
detector. Another possible way to improve the scattering efficiency of
evanescent waves is placing the sample directly in the path of an
acoustic wave, for instance, by running the wave through a
microchannel containing objects to be studied.  This approach may find
many applications in novel integrated biological/chemical detection
devices.  Finally, we comment on the possibility of extending the
proposed approach to frequencies other than the mid-IR and THz bands
discussed here.  While implementing the system for lower frequencies
is essentially trivial, near-IR and optical frequencies pose a
challenge.  Acoustic phonon energies in practical devices do not
approach the values necessary to produce a substantial wave vector
shift in these spectral bands.  However, the required shift in spatial
and temporal frequencies can in principle be attained by replacing the
acoustooptic medium with a nanostructured periodically moving grating.

In conclusion, we have proposed a system that enables detection of
sub-diffraction-limited spatial spectrum components in the far field
by utilizing scattering from an acoustic grating.  This process works
whenever the spatial frequencies of the object are comparable in scale
to the acoustic wave vector.  In its simplest implementation, the
system could aid in ``fingerprinting'' of samples based on their
subwavelength spatial features.  With the use of an additional
Bragg-shifted reference signal, it is also possible to recover the
phase of the original optical signal.  This technique, the spatial
Fourier transform infrared spectroscopy, allows to perform
subwavelength-resolved 3D imaging.  In addition, the proposed approach
has the potential to greatly enhance the specificity of mid-IR and THz
spectroscopy.

%% == end of paper:

%% Optional Materials and Methods Section
%% The Materials and Methods section header will be added automatically.

%% Enter any subheads and the Materials and Methods text below.
% \begin{materials}
\section*{Details of the calculations}
We start by considering a rectangular sound column
(i.e.\ planar acoustic wavefronts propagating in the $x$ direction)
interacting with a spectrum of incident plane waves in a dielectric
medium.  We neglect the diffraction of the sound field and assume weak
interaction.  Due to photoelastic effect, the sound field produces a
sinusoidal modulation of the dielectric permittivity~\cite{Boyd_NLO},

\begin{equation}\label{eq:epsilon_wave}
\epsilon(x) = \overline{\epsilon} + \Delta \epsilon \cos(q x-\Omega
t),\end{equation} which corresponds to a spatiotemporal volume
grating.  We may write the general form of the field inside the
grating as a sum over the discrete diffracted orders,
%Eq.~(\ref{eq:scatt_sum})
\begin{equation}\label{eq:scatt_sum}
E = \sum_j A_j(z) \exp[i(k_x + jq)x - i(\omega+j\Omega)t].
\end{equation}  The scattered plane wave components $A_j(z)$ are then governed by the Raman-Nath equations~\cite{KorpelBook},

\begin{equation}\label{eq:recursion}
A''_j(z)+k_{z_j}^2 A_j(z) = -\frac{\Delta \epsilon}{2 c^2} \omega_j^2
[A_{j-1}(z) + A_{j+1}(z)],
\end{equation}
where $k_{z_j} = \left[\overline{\epsilon}\frac{\omega^2}{c^2} - (k_x
  + j q)^2\right]^{1/2}$, and $\omega_j \simeq \omega$.

The amplitude of the $j^{\rm th}$ diffracted order, $A_j$, is
proportional to $(\Delta\epsilon)^j$, with $\Delta\epsilon/\epsilon
\ll 1$, allowing to ignore higher order terms ($j \ge 2$).  We can,
furthermore, conclude that the amount of energy scattered into the
shifted waves is small, thereby permitting to neglect the variation of
$0^{\rm th}$ diffracted order $A_0$ (the undepleted pump
approximation)~\cite{Boyd_NLO}. We note that for propagating waves,
this conclusion is valid insofar as there exists no Bragg matching
between the incident and diffracted waves.  Since the phonon wave
vector $q$ is a tunable parameter in our model, it is always possible
to pick a range of $q$ values to ensure minimal energy loss in the
incident wave. For the evanescent waves, the undepleted pump
approximation is justified by the small interaction length.

Keeping terms up to first order in Eq.~(\ref{eq:scatt_sum}), we
obtain:

\begin{equation}
A''_\pm(z)+k_{z_j}^2 A_\pm(z) = -\frac{\Delta \epsilon}{2 c^2}
\omega_j^2 A_0(z).\end{equation} The scattering amplitudes of
``upshifted'' and ``downshifted'' waves can be obtained from this
expression.  Since the input field at spatial frequencies $(k_x \mp
q)$ contributes to the output field at spatial frequency $k_x$, we may
write:
\begin{equation}\label{eq:spec_out}
E_{\rm out}(k_x) = \left[\tilde{A}_- \exp(i\Omega t) + \tilde{A}_+
  \exp(-i\Omega t) + \tilde{A}_0\right]\exp[i(k_x x - \omega
  t)],\end{equation} with $\tilde{A}_\pm \equiv t^\pm E_{\rm in}(k_x
\mp q)$, $\tilde{A}_0 \equiv t_0 A_0$. Scattering coefficients $t^\pm$
are given by

%\begin{equation}\label{eq:t0}t_0 = \frac{2
%\sqrt{\epsilon}\sqrt{(\omega/c)^2-k_x^2}}{\epsilon\sqrt{(\omega/c)^2-k_x^2}+\sqrt{\epsilon(\omega/c)^2-k_x^2}};\end{equation}
\begin{equation}\label{eq:tpm} t^\pm = t_0
\frac{\Delta\epsilon}{2}\left(\frac{\omega}{c}\right)^2\frac{1}{q(q\mp2k_x)}
\frac{ \exp(i k_z L) - \exp(i k_z^\pm L) }{ 2 k_z }(k_z+k_z^\pm),
\end{equation} where 
$k_z^\pm = \sqrt{\epsilon(\omega/c)^2-(k_x\mp q)^2}$, $k_z =
\sqrt{\epsilon(\omega/c)^2-k_x^2}$, and $t_0$ is a transmission
coefficient.  We note that for $q\approx 0$, as well as $q\approx \pm2
k_x$, the perturbative treatment of Eq.~(\ref{eq:recursion}) breaks
down due to the onset of Bragg-matching.

From Eq.~(\ref{eq:tpm}) we can estimate the diffraction efficiency of
high spatial frequency input signal $E_{\rm in}(k_x^{\rm in})$ as
\begin{equation}\label{eq:diffeff}
\left|t^\pm\right| \approx \frac{\omega/c}{2k_x^{\rm in}}
\frac{\Delta\epsilon}{n(1+n)},
\end{equation} with %$k_x^{\rm in} \approx q$, 
$n=\sqrt{\epsilon}$ (the refractive index of the acoustic medium), and
$\Delta\epsilon\propto\sqrt{F}$, the flux of acoustic energy per unit
area.

In our computations, we assume the operating wavelength of 10
\um\ with germanium as the acoustic medium.  We take
$\Delta\epsilon=10^{-3}$ and restrict the magnitude of the acoustic
wave vector $q$ to $25\,\omega/c$.  To obtain $\Delta\epsilon=10^{-3}$
the ultrasonic fluence of 33 W/cm$^2$ is required.  Since for high
spatial frequencies $k_x^{\rm in} \approx q$, acoustic driving
frequencies up to 8.75 GHz are required to retrieve $k_x^{\rm in}
\approx 25 \omega/c$.  These parameters are within reach of modern
ultrasonic transducers~\cite{HiFreqUltrasoundBulk}, as well as surface
acoustic wave devices~\cite{HiFreqUltrasoundSAW}.  %The performance of
the system depends on the characteristics of the The signals are
generated by sampling $10^6$ points in spatial frequency space.
Assuming 400 points can be sampled at once with a detector array, it
is necessary to use a focusing lens with ${\rm N.A.}\approx0.6$ and
adjust the acoustic wave vector components
$q_{x,y}\in[-25,25]\,\omega/c$ sequentially, effectively scanning the
low N.A. system over a larger spatial frequency
spectrum~\cite{BrueckOL1999,BrueckOE2007}.  At KHz readout rates,
short ($<1$ s) acquisition times can be obtained with this setup.

We simulate the measurements by taking the signal indicated by
Eqs.~(\ref{eq:measurement}) and (\ref{eq:iOut}) and introducing
additive Gaussian noise.  Assuming a shot-noise limited
long-wavelength IR detector with a typical detectivity $D^* \sim 10^6
{\rm cm \sqrt{Hz}/W}$~\cite{RoomTempDetectivityAPL} we find that
$\approx 20-30$ mW of illuminating optical power is needed to obtain
SNR of $250/k_x^{\rm in}$ used in our calculations.

Finally, we note that a potentially substantial source of noise in the
proposed system is the detection of zero-order (undiffracted)
illumination at the shifted frequencies due to the finite source
linewidth.  In the measurement region, the Lorentzian lineshape of the
source takes the same $1/\Omega$ functional dependence as the
diffracted signal of Eq.~(\ref{eq:diffeff}).  Thus, the zero-order
illumination simply adds a constant measurement background that may be
subtracted.  The noise floor is effectively raised by a factor
$1/\frac{\Omega_0}{\Delta\Omega}\frac{\Delta\epsilon}{n(1+n)}$, where
$\Omega_0=v/\lambda$ is the acoustic frequency that yields
$q=\omega/c$, and $\Delta\Omega$ is the source linewidth.  Assuming
the source is a frequency-stabilized quantum cascade laser with a
$\sim$ 15 KHz linewidth~\cite{NarrowLinewidthQCL}, this increase
factor is not significant ($\lesssim 2$).
% \end{materials}

%% Optional Appendix or Appendices
%% \appendix Appendix text...
%% or, for appendix with title, use square brackets:
%% \appendix[Appendix Title]

The work was partially supported by ARO MURI and NSF MIRTHE.

%% PNAS does not support submission of supporting .tex files such as BibTeX.
%% Instead all references must be included in the article .tex document. 
%% If you currently use BibTeX, your bibliography is formed because the 
%% command \verb+\bibliography{}+ brings the <filename>.bbl file into your

\begin{thebibliography}{10}

\bibitem{Abbe1873}
Abbe E (1873) Contributions to the theory of the microscope and microscopic detection (translated from German). \emph{Arch Mikrosk Anat} 9:413--468.

\bibitem{Lukosz1967}
Lukosz W (1967) Optical system with resolving power exceeding the classical
  limit, {II}. \emph{J Opt Soc Am} 57:932--941.

\bibitem{BrueckOL1999}
Chen X, Brueck SRJ (1999) Imaging interferometric lithography - approaching the
  resolution limits of optics. \emph{Opt Lett} 24:124--126.

\bibitem{Dragnea2001}
Dragnea B, Preusser J, Szarko JM, Leone SR, Hinsberg WDJ (2001) Pattern
  characterization of deep-ultraviolet photoresists by near-field infrared
  microscopy. \emph{J Vac Sci Technol B} 19:142--152.

\bibitem{Keilmann1999}
Knoll B, Keilmann F (1999) Near-field probing of vibrational absorption for
  chemical microscopy. \emph{Nature} 399:134--137.

\bibitem{Planken2002}
van~der Valk NCJ, Planken PCM (2002) Electro-optic detection of subwavelength
  terahertz spot sizes in the near field of a metal tip. \emph{Appl Phys Lett}
  81:1558--1560.

\bibitem{Xie1999}
Zumbusch A, Holtom GR, Xie XS (1999) Three-dimensional vibrational imaging by
  coherent anti-stokes raman scattering. \emph{Phys Rev Lett} 82:4142--4145.

\bibitem{Ito2004}
Shikata J, Matsumoto T, Zhao SL, Suzuki Y, Ito H (2004) Coherent anti-stokes
  raman spectroscopy for {THz}-frequency modes of biomolecules in aqueous
  solution. In \emph{Conference on Lasers and Electro-Optics} (Optical Society
  of America), p. CThtT36.

\bibitem{Hell2007}
Hell SW (2007) Far-field optical nanoscopy. \emph{Science} 316:1153--1158.

\bibitem{pendry}
Pendry JB (2000) Negative refraction makes a perfect lens. \emph{Phys~Rev~Lett}
  85:3966--3969.

\bibitem{PodolskiyNarimanovNSSL}
Podolskiy VA, Narimanov EE (2005) Near-sighted superlens. \emph{Opt Lett}
  30:75--77.

\bibitem{Durant2006}
Durant S, Liu Z, Steele JM, Zhang X (2006) Theory of the transmission
  properties of an optical far-field superlens for imaging beyond the
  diffraction limit. \emph{J Opt Soc Am B} 23:2383--2392.

\bibitem{LeroseyFink2007}
Lerosey G, {de Rosny} J, Tourin A, Fink M (2007) Focusing beyond the
  diffraction limit with far-field time reversal. \emph{Science}
  315:1120--1122.

\bibitem{KorpelBook}
Korpel A (1989) \emph{Acoustooptics} (Marcel Dekker, New York).

\bibitem{Boyd_NLO}
Boyd RW (2003) \emph{Nonlinear Optics} (Academic Press, San Diego), second edn.

\bibitem{OriginalGoodmanDigitalHoloPaper}
Goodman JW, Lawrence RW (1967) Digital image formation from electronically
  detected holograms. \emph{Applied Physics Letters} 11:77--79.

\bibitem{HoloCCDProblem1}
Schnars U, J\"{u}ptner W (1994) Direct recording of holograms by a ccd target
  and numerical reconstruction. \emph{Applied Optics} 33:179--181.

\bibitem{CucheDigitalHolo}
Cuche E, Marquet P, Depeursinge C (1999) Simultaneous amplitude-contrast and
  quantitative phase-contrast microscopy by numerical reconstruction of fresnel
  off-axis holograms. \emph{Appl Opt} 38:6994--7001.

\bibitem{HiFreqUltrasoundBulk}
Lanz R, Muralt P (2005) Bandpass filters for 8 {GHz} using solidly mounted bulk
  acoustic wave resonators. \emph{IEEE Transactions on Ultrasonics,
  Ferroelectrics and Frequency Control} 52:938--948.

\bibitem{HiFreqUltrasoundSAW}
Assouar MB, \emph{et~al.} (2007) High-frequency surface acoustic wave devices
  based on {A}l{N}/diamond layered structure realized using e-beam lithography.
  \emph{Journal of Applied Physics} 101:114507.

\bibitem{BrueckOE2007}
Kuznetsova Y, Neumann A, Brueck SRJ (2007) Imaging interferometric microscopy –
  approaching the linear systems limits of optical resolution. \emph{Opt
  Express} 15:6651--6663.

\bibitem{RoomTempDetectivityAPL}
Bhattacharya P, Su XH, Chakrabarti S, Ariyawansa G, Perera AGU (2005)
  Characteristics of a tunneling quantum-dot infrared photodetector operating
  at room temperature. \emph{Applied Physics Letters} 86:191106.

\bibitem{NarrowLinewidthQCL}
Williams RM, \emph{et~al.} (1999) Kilohertz linewidth from frequency-stabilized
  mid-infrared quantum cascade lasers. \emph{Opt Lett} 24:1844--1846.

\end{thebibliography}
%% .tex document. To conform to PNAS requirements, copy the reference listings
%% from your .bbl file and add them to the article .tex file, using the
%% bibliography environment described above.  

%%  Contact pnas@nas.edu if you need assistance with your
%%  bibliography.

% Sample bibliography item in PNAS format:
%% \bibitem{in-text reference} comma-separated author names up to 5,
%% for more than 5 authors use first author last name et al. (year published)
%% article title  {\it Journal Name} volume #: start page-end page.
%% ie,
% \bibitem{Neuhaus} Neuhaus J-M, Sitcher L, Meins F, Jr, Boller T (1991) 
% A short C-terminal sequence is necessary and sufficient for the
% targeting of chitinases to the plant vacuole. 
% {\it Proc Natl Acad Sci USA} 88:10362-10366.

%% Enter the largest bibliography number in the facing curly brackets
%% following \begin{thebibliography}

%%%%%%%%%%%%%%%%%%%%%%%%%%%%%%%%%%%%%%%%%%%%%%%%%%%%%%%%%%%%%%%%

%% Adding Figure and Table References
%% Be sure to add figures and tables after \end{article}
%% and before \end{document}

%% For figures, put the caption below the illustration.
%%
%% \begin{figure}
%% \caption{Almost Sharp Front}\label{afoto}
%% \end{figure}

\pagebreak

\pagebreak[4]
\clearpage
\newpage

\pagebreak[4]
\clearpage
\newpage

\pagebreak[4]
\clearpage
\newpage

%% For Tables, put caption above table
%%
%% Table caption should start with a capital letter, continue with lower case
%% and not have a period at the end
%% Using @{\vrule height ?? depth ?? width0pt} in the tabular preamble will
%% keep that much space between every line in the table.

%% \begin{table}
%% \caption{Repeat length of longer allele by age of onset class}
%% \begin{tabular}{@{\vrule height 10.5pt depth4pt  width0pt}lrcccc}
%% table text
%% \end{tabular}
%% \end{table}

%% For two column figures and tables, use the following:
%% \begin{figure*}
%% \caption{Almost Sharp Front}\label{afoto}
%% \end{figure*}

%% \begin{table*}
%% \caption{Repeat length of longer allele by age of onset class}
%% \begin{tabular}{ccc}
%% table text
%% \end{tabular}
%% \end{table*}

\end{document}